\documentstyle[12pt]{article}
\def\a{\alpha}
\def\b{\beta}

\def\d{\delta}

\def\p{\psi}

\def\L{\Lambda}

\font\sqi=cmssq8
\def\DC{\kern2pt {\hbox{\sqi I}}\kern-4.2pt\rm C}

\def\be{\begin{equation}}
\def\ee{\end{equation}}
\def\arr{\begin{array}{rll}}
\def\ea{\end{array}}
\def\bea{\begin{eqnarray}}
\def\eea{\end{eqnarray}}

\begin{document}
%\large
\renewcommand{\thefootnote}{\fnsymbol{footnote}}
\begin{titlepage}
\noindent

%\vskip 3.0cm

\begin{center}

{\Large\bf New insight into WDVV equation}\\
\bigskip
%{\Large\bf Calogero model }\\
\bigskip

\vskip 0.5cm

{\large S. Bellucci}${}^a$\footnote{bellucci@lnf.infn.it},
{\large A.V. Galajinsky}${}^b$\footnote{galajin@mph.phtd.tpu.edu.ru},
{\large E. Latini}${}^a$\footnote{latini@lnf.infn.it}

\vskip 0.2cm

${}^a$ {\it INFN--Laboratori Nazionali di Frascati, C.P. 13,
00044 Frascati, Italy}\\

\vskip 0.2cm
${}^b${\it Laboratory of Mathematical Physics, Tomsk Polytechnic University, \\
634050 Tomsk, Lenin Ave. 30, Russian Federation}

\vskip 0.2cm

\end{center}
\vskip 1cm

\begin{abstract}
We show that Witten-Dijkgraaf-Verlinde-Verlinde equation underlies the construction of
$N=4$ superconformal multi--particle mechanics in one dimension, including a
$N=4$ superconformal Calogero model.
\end{abstract}

\vspace{0.5cm}


\end{titlepage}
\renewcommand{\thefootnote}{\arabic{footnote}}
\setcounter{footnote}0
\noindent
{\bf 1. Introduction}\\

Witten-Dijkgraaf-Verlinde-Verlinde (WDVV) equation was originally formulated in the context of $2d$ topological
field theory \cite{w,dvv}. Being a consequence of specific recursion relations for correlation functions
which underlie a $2d$ topological model \cite{w}, the WDVV equation can be understood as providing a
ring structure for local observables of a perturbed topological theory \cite{dvv,dvv1}.
As within the topological framework the identity operator is physical, one coordinate
in a parameter space of a perturbed topological model plays a distinguished role. In particular, some of the
WDVV coefficients are identified with
the two--point correlator on a sphere, and the latter is assumed to be a constant nondegenerate matrix
(a metric on a physical Hilbert space) \cite{dvv,dvv1}. A very attractive link between the WDVV equation and
differential geometry was established in ref. \cite{dubr}
where it was demonstrated that locally any solution of the WDVV equation defines in the domain the structure
of a Frobenius manifold and vice versa.

A generalization of the WDVV equation which treats all the coordinates on equal footing was proposed in
ref. \cite{itep} (see also ref. \cite{bon}).  The corresponding
solution proved to be applicable to Seiberg--Witten theory where it was
interpreted  as the prepotential entering the low--energy effective action of
$N=2$ supersymmetric Yang--Mills theory \cite{itep}.

The purpose of the present paper is to discuss one more physical
application of the WDVV equation. As will be demonstrated below,
in a slightly modified form it underlies the construction of $N=4$
superconformal multi--particle models in one dimension, including
a $N=4$ superconformal Calogero model.

Notice also that there is an extensive literature on $N=4$
supersymmetric mechanics in various dimensions. So far, the
attention focused mainly on
building appropriate Lagrangian and Hamiltonian models of
supersymmetric mechanics and studying partial supersymmetry
breaking. For approaches similar to ours see refs.
\cite{don,don1}, which report on $N=4$ supersymmetric mechanics in arbitrary $D$ and
also contain an extensive list of references on $N=4$ supersymmetric
mechanics. In \cite{bp} $N=4, D=2$ supersymmetric mechanics described
by chiral superfield actions was considered. Geometric models of $N=4$ supersymmetric mechanics
with $(2d|2d)_{\DC}$ dimensional phase superspace have been proposed in \cite{bn}. These models
reduce to the one considered in \cite{bp}, in the simplest case of $d=1$ and no central charges.

The interest in $N=4$ superconformal multi--particle mechanics is due in part to a potential application
in black hole physics. It was conjectured by Gibbons and Townsend \cite{gib} that
the large $n$ limit of a $N=4$ superconformal Calogero model might be relevant for a microscopic description
of the extreme Reissner--Nordstr\"om black hole, at least near the horizon. The conjecture originated
from the fact that the near horizon geometry in this case has the isometry group $SU(1,1|2)$ which
includes the conformal group $SO(1,2)$ as a subgroup. Taking into account that
the extreme Reissner--Nordstr\"om black hole can be viewed as the configuration of four intersecting
supergravity $D3$--branes wrapped on $T^6$~\cite{kleb,bal} and assuming that each of the supergravity
$D3$--branes can be interpreted as a large number of coinciding microscopic $D3$--branes~\cite{gib},
one comes to the conjecture that there exists a $SU(1,1|2)$--invariant mechanics governing the
fluctuation of the branes in the region of intersection~\cite{gib}. Being conformally invariant, the
Calogero model seems to be a good candidate for the bosonic limit of such a $N=4$ superconformal multi--particle
mechanics.

Motivated by these issues in the next section we construct a set of generators yielding a
representation of $su(1,1|2)$ superalgebra on a phase space which includes $n$ copies of
that corresponding to a conventional $N=4$ superconformal particle \cite{ikl}. These are designed to describe
symmetries of a multi--particle mechanics in one dimension. As the Hamiltonian is part of the superalgebra,
the dynamics is automatically taken into account. One reveals a central charge in the superalgebra
which shows up in the Poisson brackets of the supersymmetry charges with the superconformal ones.

The explicit form of the generators involves two scalar functions $V$ and $F$ which depend on position
of particles only. The first of them specifies a potential of the problem, while the second function
controls higher order
fermionic contributions into the Hamiltonian. The structure relations of $N=4$ superconformal algebra force
$V$ and $F$ to obey a system of partial differential equations. One of them states that the vector
$\partial^i V$ is covariantly constant with respect to the "connection"
$\partial^i \partial^j \partial^k F \equiv W^{ijk}$. Then
the WDVV equation, $W^{ijk} W^{kpl}=W^{ljk} W^{kpi}$, arises as the integrability
condition. Notice that, similarly to the case of topological theories, the metric entering the equation
is flat (the indices are contracted with the use of the ordinary Euclidean metric).
However, there is no distinguished coordinate around and, as thus, no identification of some of the WDVV
coefficients with the metric hold.

As we discuss in more detail below, the arising system of partial differential equations can be treated in two
ways. One can take a particular solution of the WDVV equation (the prepotential F) and then construct a $N=4$
superconformal multi--particle model associated to it (the prepotential V). Alternatively, one can start with a
bosonic conformally invariant multi--particle mechanics and then look for a solution of the
WDVV equation that will provide a $N=4$ superconformal extension.

Motivated by the search for a $N=4$ superconformal Calogero model, in this paper we follow the second line.
In section 3 we take $V$ which reproduces the Calogero potential and then move on to fix $F$. As will be shown below,
in general, the solution is not unique. We treat in detail the two--body problem and show that at the classical level
there are two families of $N=4$ superconformal models which are parameterized by the value of the
central charge
in the $su(1,1|2)$ superalgebra. We present also partial results on the three--body problem. In particular,
the nonlinear WDVV equations are reduced to a system of four linear partial differential equations of the Euler
type. In section 4 we summarize our results and discuss possible further
developments.
\\

\noindent
{\bf 2. Algebraic structure}\\

First of all, let us fix the notation. Working in phase space, different particles are labeled by the
index $i=1,\dots,n$ attached to a pair $(x,p)$. These variables obey the conventional Poisson bracket
\be
\{x^i,p^j \}=\d^{ij}.
\ee
Although we are primarily concerned with the construction of $N=4$ superconformal many--body models,
it seems natural to require that the conventional $N=4$ superconformal
mechanics (see ref. \cite{ikl} for details) be reproduced in the one--particle case. In particular,
this suggests the correct way to introduce fermions. One assigns a $SU(2)$
spinor $\p_\a$, $\a=1,2$ and its complex
conjugate to each particle. Our convention for the conjugation reads
\be
{(\p^i_\a)}^{+}={\bar\p}^{i \a}.
\ee

In what follows we use the bracket
\be
\{\p^i_\a,{\bar\p}^{j \b} \}=-i  {\d_\a}^\b \d^{ij}
\ee
and sum up over repeated indices.\\

\noindent
{\it 2.1 $R$--symmetry generators}\\

In order to construct a representation of $su(1,1|2)$ superalgebra
on the phase space described above, it proves to be instructive to
start with $R$--symmetry generators. For the case at hand, they
form $su(2)$ subalgebra. Given the latter and half of the
supercharges, one can consistently fix the other half. The same
applies to superconformal generators. Besides, having chosen the
$su(2)$ currents, one could also specify the dependence of the odd
generators on the fermionic coordinates.

In this paper we adhere to the simplest representation of $su(2)$ which can be constructed out of
the fermionic variables alone
\bea\label{su2}
&&
J_{+}=-i\psi^i_1  \p^i_2, \qquad
J_{-}=-i {\bar\psi}^{i 1} {\bar\psi}^{i 2}, \qquad
J_3=\textstyle{\frac{1}{2}}(\p^i_1 {\bar\p}^{i 1} + \p^i_2 {\bar\p}^{i 2}).
\eea
These generators obey the standard algebra
\be
\{J_{+},J_{-}\}=-2i J_3, \quad \{J_{+},J_3\}=iJ_{+}, \quad \{J_{-},J_3\}=-iJ_{-}.
\ee
Notice that $J_3$ measures the $U(1)$ charge of the fermionic
coordinates
\be
\{ \p^i_\a,J_3 \}=\textstyle{\frac{i}{2}} \p^i_\a, \qquad \{ {\bar\p}^{i \a}, J_3 \}=-\textstyle{\frac{i}{2}}
{\bar\p}^{i \a}.
\ee
It is worth mentioning also that $J_{+}$ and $J_{-}$ are complex conjugates of each other, while $J_3$ is a
self--conjugate generator.\\

\noindent
{\it 2.2 Supersymmetry charges}\\

We now proceed to the construction of four supersymmetry charges
$G_1$, $G_2$, ${\bar G}^1$, ${\bar G}^2$. Taking into account the brackets
\bea\label{rot}
&&
\{G_1,J_{+}\}=0, \qquad \{G_1,J_{-}\}={\bar G}^2, \qquad
\{G_1,J_3\}=\frac{i}{2} G_1,
\nonumber\\[2pt]
&&
\{G_2,J_{+}\}=0, \qquad \{G_2,J_{-}\}=-{\bar G}^1, \qquad \{G_2,J_3\}=\frac{i}{2}G_2,
\eea
which make part of $N=4$ superconformal algebra in one dimension,
and the conjugation properties
\be\label{str}
{(G_1)}^{+}={\bar G}^1, \qquad {(G_2)}^{+}={\bar G}^2,
\ee
one concludes that it is instructive to start with an Ansatz for $G_1$. Then, the rest can be fixed with the use
of eq. (\ref{rot}), the $N=4$ supersymmetry algebra
\be\label{ggh}
\{G_\a, {\bar G}^\b \}=-2iH {\d_\a}^\b
\ee
and the conjugation relations. The function $H$ appearing in the r.h.s. defines the Hamiltonian of a
multi--particle system we aim at.

As it was mentioned earlier, the one--particle case should
reproduce the conventional $N=4$ superconformal mechanics. Since
the supersymmetry generators of the latter are cubic in the
fermionic coordinates, for a generic multi--particle model it
seems reasonable to terminate the expansion in fermions at that
order. Then, a straightforward calculation yields (${\bar G}^1$
and ${\bar G}^2$ follow by hermitian conjugation)
\bea\label{superg} && G_1=(p^i+i \partial^i V(x)) \p^i_1 +(p^i-i
\partial^i V(x)) \p^i_2 -i W^{ijk} (x) \p^i_1 \p^j_2 {\bar \p}^{k
1} +i W^{ijk} (x) \p^i_1 \p^j_2 {\bar \p}^{k 2},
\nonumber\\[2pt]
&&
G_2=(p^i+i \partial^i V(x)) \p^i_1 -(p^i-i \partial^i V(x)) \p^i_2 +i W^{ijk} (x) \p^i_1 \p^j_2 {\bar \p}^{k 1}
+ i W^{ijk} (x) \p^i_1 \p^j_2 {\bar \p}^{k 2},
\nonumber\\[2pt]
&&
\eea
%\bea
%\nonumber\\[2pt]
%&&
%{\bar G}^1=(p^i-i \partial^i V(x)) {\bar\p}^{i1}+\a (p^i+i \partial^i V(x)) {\bar\p}^{i2}
%-\a i W^{ijk} (x) \p^i_1 {\bar\p}^{j1} {\bar\p}^{k2}+iW^{ijk} (x) \p^i_2 {\bar\p}^{j1} {\bar \p}^{k2},
%\nonumber\\[2pt]
%&&
%{\bar G}^2 =\a (p^i-i \partial^i V(x)) {\bar\p}^{i1}-(p^i+i \partial^i V(x)) {\bar\p}^{i2}
%+\a i W^{ijk} (x) \p^i_2 {\bar\p}^{j1} {\bar\p}^{k2}+iW^{ijk} (x) \p^i_1 {\bar\p}^{j1} {\bar \p}^{k2}.
%\nonumber\\[2pt]
%&&
%\eea
where the coefficient functions $V(x)$ and $W^{ijk}(x)$ obey the following restrictions:
\be\label{pot}
\partial^i \partial^j V-W^{ijk} \partial^k V=0,
\ee \be\label{wdvv} W^{ijk}=\partial^i \partial^j  \partial^k F,
\qquad W^{ijk} W^{kpl}=W^{ljk} W^{kpi}, \ee where $F$ is a scalar.
Surprisingly enough, in the last line one recognizes the
Witten-Dijkgraaf-Verlinde-Verlinde equation which underlies a $2d$
topological field theory \cite{w,dvv}.

In the context
of the Calogero model considerations like the above ones were
presented in ref. \cite{wyl}, where a similar set of equations was
derived. However, only partial solutions for $V$ and $F$ were
examined which led the author to conclude that there was no a
$SU(1,1|2)$--invariant extension of the Calogero model for generic
values of coupling constant. Below we give a more rigorous
procedure of solving these equations which relies upon general
solutions for $V$ and $F$.

According to eq. (\ref{ggh}) the Hamiltonian which governs the dynamics of the model has the form
\bea\label{ham}
&&
H=p^i p^i+\partial^i V(x) \partial^i V(x)-2\partial^i \partial^j V(x) \p^i_1 {\bar\p}^{j 1}
+2\partial^i \partial^j V(x) \p^i_2 {\bar\p}^{j 2} +2\partial^i W^{jkl} (x) \p^i_1 \p^j_2 {\bar\p}^{k 1}
{\bar\p}^{l 2}.
\nonumber\\[2pt]
&&
\eea

Thus, given a solution of the WDVV equation (\ref{wdvv}), one can construct a $N=4$
supersymmetric multi--particle mechanics associated to it. The bosonic part of the corresponding
potential is of the form $\partial^i V \partial^i V$. It should be mentioned that
eqs. (\ref{pot}),(\ref{wdvv}) hold also if one examines a smaller $N=2$ subalgebra.

The inverse problem is also worthy to consider. One can start with a model governed by a potential
$U(x^1,\dots, x^n)$
and set
\be
U(x)=\partial^i V (x) \partial^i V (x),
\ee
which will fix $V$. In general, the corresponding solution is not unique.
Then, solving the system of partial differential equations (\ref{pot}),(\ref{wdvv}) for $W^{ijk}$
will provide a $N=4$ supersymmetric extension of the model we started with. Below, we examine
this possibility for the Calogero model.

Curiously enough, eqs. (\ref{pot}),(\ref{wdvv}) can be restated as the condition of the existence of a
vector field $V^i (x)$ covariantly constant with respect to the "connection"
$W^{ijk}=\partial^i \partial^j \partial^k F$
\be
{\mathcal{D}}^i V^j=\partial^i V^j-W^{ijk} V^k=0.
\ee
As $W^{ijk}$ is completely symmetric, the previous line implies
\be
\partial^i V^j-\partial^j V^i=0 \quad \Longrightarrow \quad V^i=\partial^i V,
\ee
for some scalar $V$, while the WDVV equation comes about as the integrability condition
\be
[ {\mathcal{D}}^i, {\mathcal{D}}^j ]V^k=0.
\ee
It would be interesting to find a geometrical interpretation behind these equations. Because the problem is
formulated in a flat space, we anticipate that a geometric description, if any, is related to symplectic structures on
the manifold (in this respect see refs. \cite{nut,nut1}).

Concluding this section we notice that the $su(2)$ generators displayed in eq.
 (\ref{su2}) commute with the
Hamiltonian (\ref{ham}). Conservation of the supersymmetry charges follows from eq. (\ref{ggh})
and Jacobi identities.\\

\noindent
{\it 2.3 Superconformal generators}\\

Now let us see what restrictions on the form of the functions $V$ and $W^{ijk}$
follow if one wishes to make the model superconformal. Guided by the experience with
the conformal mechanics \cite{aff}, one takes the generators of dilatations and special conformal transformations in the form
\be\label{dil}
D=tH-\frac{1}{2}x^i p^i, \quad K=t^2 H -t x^i p^i +\frac{1}{2} x^i x^i.
\ee
Being combined with the Hamiltonian they obey the conformal algebra in one dimension
\be
\{H,D \}=H, \quad \{H,K \}=2D, \quad \{D,K \}=K,
\ee
provided
%\be
%x^i \partial^i \partial^j V+\partial^j V=0, \qquad x^i \partial^i W^{jkl}+W^{jkl}=0.
%\ee
%In other words
\be\label{c1}
x^i \partial^i V=C_1, \qquad x^i W^{ijk}=C_2 \d^{jk},
\ee
where $C_1$ and $C_2$ are arbitrary real constants.

At this stage only superconformal generators, which in what follows we denote by $S_\a$ and ${\bar S}^\a$, are missing.
Taking into account that the structure relations of $su(1,1|2)$ superalgebra involve the bracket
\be
\{ K, G_\a \}=-S_\a,
%\qquad \{K, {\bar G}^\a \}=-{\bar S}^\a,
\ee
these can be consistently specified
\bea
&&
S_1=tG_1-\frac{1}{2} x^i \p^i_1
-\frac{1}{2} x^i \p^i_2, \qquad
S_2=tG_2-\frac{1}{2} x^i \p^i_1
+\frac{1}{2} x^i \p^i_2.
\eea
The generators ${\bar S}^1$ and ${\bar S}^2$ follow by hermitian conjugation.

Finally, it remains to check that the entire algebra closes. This proves to be the case
provided
\be
C_2=-1.
\ee
The last constraint follows immediately if one considers Poisson brackets of the superconformal generators with
the supersymmetry ones. The explicit verification of the algebra turns out to be rather tedious. The use of
Jacobi identities considerably simplifies the calculation. In the basis chosen one finds
(vanishing brackets are omitted)
\bea
&&
\{D,G_\a\}=-\frac{1}{2} G_\a, \quad
\{D,S_\a\}=\frac{1}{2} S_\a, \quad
\{H,S_\a\}=G_\a, \quad \{S_\a,{\bar S}^\b\}=-2iK {\d_\a}^\b,
\nonumber\\[2pt]
&&
\{S_1,J_{-}\}={\bar S}^2,\quad
\{S_1,J_3\}=\frac{i}{2} S_1, \quad \{S_2,J_{-}\}=-{\bar S}^1,\quad
\{S_2,J_3\}=\frac{i}{2} S_2,
\nonumber\\[2pt]
&&
\{G_1,S_2\}=-2iJ_{+}, \quad \{G_1,{\bar S}^1\}=-2iD+2J_3, \quad
\{G_1,{\bar S}^2\}=Z,
\nonumber\\[2pt]
&&
\{G_2,S_1\}=2iJ_{+},
 \quad \{G_2,{\bar S}^1\}=Z,
\quad \{G_2,{\bar S}^2\}=-2iD+2J_3,
\eea
plus their complex conjugates. Notice the appearance of the central charge
\be
Z=-C_1
\ee
with $C_1$ from eq. (\ref{c1}), in the brackets involving the supersymmetry generators and the superconformal ones.

Let us recapitulate. On a phase space spanned by $2n$ real bosons $(x^i,p^i)$ and
$4n$ real fermions $(\p^i_\a, {\bar\p}^{i \a})$, $i=1,\dots,n$, $\a=1,2$, one can build a representation of
$N=4$ superconformal algebra. The corresponding generators contain two scalar
functions $V$ and $F$ which obey the system of partial differential equations
\bea\label{final}
&&
x^i \partial^i V=C_1, \qquad \partial^i \partial^j V-W^{ijk} \partial^k V=0,
\eea
\bea\label{final-1}
&&
x^i W^{ijk}+\d^{jk}=0, \qquad W^{ijk} W^{kpl}=W^{ljk} W^{kpi},
\eea
where $W^{ijk}=\partial^i \partial^j  \partial^k F$ and $C_1$ is an arbitrary constant.
The bosonic part of the Hamiltonian which governs the dynamics of the resulting N=4 superconformal many--body
model reads
\be
H_{Bose}=p^i p^i+\partial^i V \partial^i V.
\ee
It is worth mentioning that no contradiction results from contracting eqs. (\ref{final}),(\ref{final-1}) with
$x^i$, $\partial^i V$, or $W^{ijk}$. This provides a naive compatibility check for the system.\\

\noindent
{\bf 3. $N=4$ superconformal Calogero model}\\

As it was mentioned earlier, there are two ways to treat the system (\ref{final}),(\ref{final-1}). One can start
with a particular solution of the WDVV equation and then construct a $N=4$
superconformal multi--particle model associated to it. Alternatively, one can take a specific bosonic
conformally invariant multi--particle mechanics and then search for a solution of the WDVV equation that
will provide a $N=4$ superconformal extension of the theory. For the rest of the paper we take the second path
and consider the Calogero model which is governed by the Hamiltonian \cite{cal}
\be\label{calogero}
H=p^i p^i+{\sum}_{i<j}~ \frac{g^2}{{(x^i-x^j)}^2}.
\ee
Here $g$ is a dimensionless coupling constant. It was known for a long time that the Calogero model exhibits
the conformal invariance \cite{reg,woj}. Yet a consistent $N=4$ superconformal extension has not been constructed.
For earlier work on this subject, see refs. \cite{wyl,ghosh,gal,bgk}.
\\

\noindent
{\it 3.1 One--particle case}\\

For the one particle case the system (\ref{final}),(\ref{final-1})
reduces to ordinary differential equations which are easily integrated
\be
V=\pm g \ln |x| +C_0, \qquad W=-\frac{1}{x}, \qquad F=-\frac{1}{2} x^2 \ln |x|.
\ee
with $C_0$ a constant. Notice that by definition $F$ is defined up to a polynomial quadratic in $x$. In what follow
we disregard those terms as they do not contribute to $W$.  The WDVV equation is trivial.
The corresponding $N=4$ superconformal mechanics coincides with that constructed in ref. \cite{ikl}. \\

\noindent
{\it 3.2 Two--body problem}\\

Let us start with the linear inhomogeneous partial differential equation $x^i \partial^i V=C_1$, $i=1,2$.
Making the substitution $V=C_1 \ln |x^1| +\tilde V $
one reduces it to the homogeneous equation $x^i \partial^i \tilde V=0$ which is equivalent to the system of ordinary differential
equations $\frac{d x^1}{x^1}=\frac{d x^2}{x^2}$.
The only integral of the latter, $\frac{x^1}{x^2}=const$, provides
the general solution $\tilde V=\L (\textstyle{\frac{x^1}{x^2}})$ for the former,
with $\L$ being an arbitrary function. As a result, for $V$ one has
\be
V=C_1 \ln |x^1| +\L (\textstyle{\frac{x^1}{x^2}}).
\ee

We next wonder what $\L$ corresponds to the Calogero potential. This information is encoded in the
condition
\be
{(\partial^1 V)}^2 +{(\partial^2 V)}^2=\frac{g^2}{{(x^1-x^2)}^2},
\ee
which causes $\L$ to obey the ordinary differential equation
\be
(1+y^2) y^2 {\left(\frac{d\L}{d y} \right)}^2 +2 y C_1 \frac{d\L}{d y} +C_1^2 -\frac{g^2 y^2}{{(y-1)}^2}=0,
\ee
where we denoted $\frac{x^1}{x^2}=y$. This is a quadratic algebraic equation
with respect to the derivative which means that there are two distinct possibilities for $\L$.
The solutions explicitly depend on the constant $C_1$ and acquire the simplest form for the following
choice:
\be\label{c_1}
{\left(\frac{C_1}{g}\right)}^2=\frac{1}{2}.
\ee
For this particular value one finds
\be
\L_1=C_1 \ln |1-\frac{1}{y}|+const,
\ee
and
\be
\L_2=C_1 \ln |\frac{y^2+1}{y(y-1)}|+const.
\ee

Thus, given the two--body Calogero model, the prepotential $V$ generating a $N=4$ superconformal extension is not
unique. In fact, there are two families of models which are parameterized by the central charge $C_1$.
For example, for the
particular value of $C_1$ displayed in eq. (\ref{c_1}) above one gets either
\be\label{v1}
V_1=C_1 \ln |x^1-x^2|+const,
\ee
which is of common use in the literature or
\be\label{v2}
V_2=C_1 \ln |\frac{{(x^1)}^2 +{(x^2)}^2}{x^1-x^2}|+const,
\ee
which is an alternative.
When constructing a $N=4$ superconformal extension of the two-body problem both options
are to be considered on equal footing. Worth mentioning also is that both solutions exhibit cyclic symmetry.

Having fixed $V$, let us discuss the equations which determine $F$. Taking into account that in the context of
our problem $F$ is defined up to a second order polynomial in the variable $x$, one can bring the first equation in
(\ref{final-1}) to the form
\be\label{f}
x^i \partial^i F -2 F + \frac{x^i x^i}{2}=0,
\ee
which is an analog of the scaling law for the free energy of a perturbed topological theory
(see ref. \cite{dvv1} for more details). This is an inhomogeneous partial differential equation which
is readily integrated by conventional means (see e.g. ref. \cite{smirnov}). The general solution reads
\be
F=-\frac{1}{2} x^i x^i \ln |x^1| +\frac{1}{2} {(x^1)}^2 \Delta(\textstyle{\frac{x^1}{x^2}}) ,
\ee
where $\Delta$ is an arbitrary function of the ratio $\textstyle{\frac{x^1}{x^2}}$.

With $F$ at hand one can calculate the WDVV coefficients. These take the most readable form
if one introduces into consideration a subsidiary function
\be\label{sig}
\Sigma (y)=\frac{1}{2} y^3 \frac{d^3 \Delta}{d y^3} +3 y^2 \frac{d^2 \Delta}{ d y^2} +
3 y\frac{d \Delta}{d y},
\ee
where as above we denoted $\frac{x^1}{x^2}=y$. Making use of the latter one derives the following expressions for
$W^{ijk}$:
\bea\label{w-coef}
&&
W^{111}=-\frac{1}{x^1} (1+ \frac{1}{y^2} -\Sigma (y)), \qquad
W^{112}=\frac{1}{x^1} (\frac{1}{y}- y \Sigma (y) ),
\nonumber\\[2pt]
&&
W^{122}=-\frac{1}{x^1} (1-y^2 \Sigma(y)), \qquad W^{222}=-\frac{1}{x^1} y^3 \Sigma(y).
\eea
The convenience of such a representation becomes evident if one turns to analyze
the WDVV equation. A simple calculation shows that the only nontrivial condition
for the two--particle case
\be
W^{11k} W^{k22}=W^{12k} W^{k12},
\ee
is identically satisfied without imposing any restriction on the form of the function $\Delta(y)$.

Thus, it remains to solve the second equation in (\ref{final}) which involves both $V$ and $F$.
No special effort is needed to demonstrate that three components forming the system are functionally
dependent as there are two identities between them. This statement is easily verified with the use of
the condition $x^i\partial^i V=C_1 $. The only independent equation -- the $"22"$--component -- allows one to
express $\Sigma$ in terms of $V$ algebraically
\be
\Sigma=\frac{\partial^1 V + x^1 \partial^2 \partial^2 V}{y^2 (\partial^1 V -y \partial^2 V)}.
\ee

Taking into account the explicit form of $\Sigma$ from eq. (\ref{sig})
one finally arrives at the inhomogeneous Euler equation
which completely determines $\Delta$ and, as thus, $F$.

As we have seen above, given the value of $C_1$, there are
two prepotentials $V_1$ and $V_2$ which reproduce the two--particle Calogero model in the bosonic limit.
As an example, consider $V_1$ and $V_2$ displayed in eqs. (\ref{v1}),(\ref{v2}) above. For this particular choice
of $C_1$ one readily gets
\be
\Sigma_1(y)=\frac{1}{y^2(1-y^2)},
\ee
and
\be
\Sigma_2(y)=\frac{1}{y^2(1-y^2)}+\frac{4y(1-y)}{(1+y) {(1+y^2)}^2},
\ee
respectively. In view of eq. (\ref{w-coef}) the WDVV coefficients
can be calculated by purely algebraic means. Thus, for the two--body problem actually there is no need to
know the explicit form of the prepotential $F$.\\

\noindent
{\it 3.3 Three--body problem}\\

In the preceding section we have constructed a $N=4$ superconformal extension for the two--particle
Calogero model following the recipe proposed in section 2. However, that the extension
exists might have been expected on general grounds. Indeed, employing the canonical transformation
\bea
&&
(x^1,p^1) \quad \longrightarrow \quad (\frac{1}{\sqrt{2}}(x^1-x^2),\frac{1}{\sqrt{2}}(p^1-p^2)),
\nonumber\\[2pt]
&&
(x^2,p^2)  \quad \longrightarrow \quad (\frac{1}{\sqrt{2}}(x^1+x^2),\frac{1}{\sqrt{2}}(p^1+p^2)),
\eea
one brings the Hamiltonian of the two--body Calogero model to the sum of a free particle (the center of mass)
and the conventional conformal mechanics \cite{aff}, both admitting a $N=4$ superconformal
extension. Thus, the most interesting situation occurs when the number of particles is greater
than two. In this section we discuss the three--body problem.

For the case of three particles the system of ordinary differential equations corresponding to the
homogeneous equation $x^i \partial^i \tilde V=0$ has two independent integrals
\be
\frac{x^1}{x^2}=const, \qquad \frac{x^1}{x^3}=const'.
\ee
As a result, the general solution of the first equation in (\ref{final}) reads
\be
V=C_1 \ln |x^1| +\L (y,z),
\ee
where $\L$ is an arbitrary function of two variables $y=\frac{x^1}{x^2}$ and
$z=\frac{x^1}{x^3}$. In order to pick out $\L$ which corresponds to the Calogero model
one has to impose the constraint
\be
{(\partial^1 V)}^2 +{(\partial^2 V)}^2 +{(\partial^3 V)}^2= \frac{g^2}{{(x^1-x^2)}^2}+\frac{g^2}{{(x^1-x^3)}^2}+
\frac{g^2}{{(x^2-x^3)}^2},
\ee
which implies a partial differential equation for $\L$. Like the two--body case there are
many solutions to the latter equation. The simplest one reads
\be
V=\frac{C_1}{3} \left(\ln |x^1-x^2|+\ln|x^1-x^3|+\ln |x^2-x^3| \right),
\ee
where ${(C_1)}^2={(3g)}^2/2$.

In much the same way the general solution of eq. (\ref{f}) is found to be of the form
\be
F=-\frac{1}{2} x^i x^i \ln |x^1| +\frac{1}{2} {(x^1)}^2 \Delta(y,z),
\ee
where $\Delta(y,z)$ is an arbitrary function. Triple differentiation of $F$ then gives ten WDVV coefficients
\bea\label{wdvv-c}
&&
W^{111}=-\frac{1}{x^1} (1 +\frac{1}{y^2} +\frac{1}{z^2} -\Sigma_1 -\Sigma_2 -3\Sigma_3
-3\Sigma_4),
\nonumber\\[2pt]
&&
W^{112}=\frac{1}{x^1} (\frac{1}{y} -y  \Sigma_1 -y \Sigma_3 -2y \Sigma_4),
\quad W^{113}=\frac{1}{x^1}
(\frac{1}{z} -z \Sigma_2 -2 z \Sigma_3 - z \Sigma_4),
\nonumber\\[2pt]
&&
W^{122}=-\frac{1}{x^1}(1-y^2 \Sigma_1 - y^2 \Sigma_4), \quad
W^{123}=\frac{1}{x^1}y z( \Sigma_3 + \Sigma_4),
\nonumber\\[2pt]
&&
W^{133}=-\frac{1}{x^1}(1-z^2 \Sigma_2 - z^2 \Sigma_3), \quad
W^{222}=-\frac{1}{x^1} y^3 \Sigma_1, \quad W^{223}=-\frac{1}{x^1} z y^2 \Sigma_4,
\nonumber\\[2pt]
&&
W^{233}=-\frac{1}{x^1} y z^2 \Sigma_3, \quad W^{333}=-\frac{1}{x^1} z^3 \Sigma_2,
\eea
which are conveniently expressed in terms of four subsidiary functions
\bea\label{subs}
&&
\Sigma_1 =\frac{1}{2} y^3 \frac{{\partial}^3 \Delta}{\partial y^3} +3 y^2 \frac{{\partial}^2 \Delta}{\partial y^2}
+3y \frac{\partial \Delta}{\partial y}, \qquad
%\nonumber\\[2pt]
%&&
\Sigma_2 =\frac{1}{2} z^3 \frac{{\partial}^3 \Delta}{\partial z^3} +3 z^2 \frac{{\partial}^2 \Delta}{\partial z^2}
+3z \frac{\partial \Delta}{\partial z},
\nonumber\\[2pt]
&&
\Sigma_3=\frac{1}{2} y z^2 \frac{{\partial}^3 \Delta}{\partial y \partial z^2} +
y z \frac{{\partial}^2 \Delta}{\partial y \partial z},  \qquad \qquad \quad
\Sigma_4=\frac{1}{2} z y^2 \frac{{\partial}^3 \Delta}{\partial z \partial y^2} +
y z\frac{{\partial}^2 \Delta}{\partial y \partial z}.
\eea

As we see form eq. (\ref{wdvv-c}) the first six functions can be expressed in terms of the latter four.
This observation simplifies an analysis of the WDVV equations considerably. Actually, it is a matter
of straightforward calculation to verify that out of six WDVV equations which are present for the case
under consideration
\bea
&&
W^{11k} W^{k22}=W^{12k}W^{k12}, \quad W^{11k} W^{k23}=W^{12k}W^{k13}, \quad W^{11k} W^{k33}=W^{13k}W^{k13},
\nonumber\\[2pt]
&&
W^{12k} W^{k23}=W^{13k}W^{k22}, \quad W^{12k} W^{k33}=W^{13k}W^{k23}, \quad W^{22k} W^{k33}=W^{23k}W^{k23},
\eea
only the last one is independent while the first five equations contain just the same information.

Thus, the only constraint on the model coming on the WDVV side reads
\be\label{w-f}
W^{22k} W^{k33}=W^{23k}W^{k23},
\ee
which can be viewed as an algebraic equation involving four subsidiary functions
$\Sigma_1$, $\Sigma_2$, $\Sigma_3$, $\Sigma_4$.

As the last step we analyze the second equation in (\ref{final}). Similar to the two--body problem one
can demonstrate that among six relations available for this case only half, namely the $"22"$--, $"23"$--, and
$"33"$--components, are independent.
%\bea
%&&
%x^1 \partial^2 \partial^2 V +\partial^1 V (1-y^3 \Sigma_1 -z y^3 \Sigma_4)  +\partial^2 V y^4 \Sigma_1
%+\partial^3 V z^2 y^3 \Sigma_4 =0,
%\nonumber\\[2pt]
%&&
%x^1 \partial^2 \partial^3 V-\partial^1V y^2 z^2 (\Sigma_3+\Sigma_4)+\partial^2 V z^2 y^3 \Sigma_4
%+\partial^3 V y^2 z^3 \Sigma_3=0,
%\nonumber\\[2pt]
%&&
%x^1 \partial^3 \partial^3 V+\partial^1 V (1-z^3\Sigma_2 -yz^3\Sigma_3)+\partial^2 V y^2 z^3 \Sigma_3+
%\partial^3 V z^4 \Sigma_2=0.
%\eea
These are simple algebraic equations which allow one to express $\Sigma_1$, $\Sigma_2$ and $\Sigma_3$ in
terms of $V$ and $\Sigma_4$. Substituting the resulting expressions in eq. (\ref{w-f}) one finally gets
a quadratic algebraic equation which completely determines $\Sigma_4$ in terms of $V$.

Thus, similar to the two--body case, all the subsidiary functions are fixed by purely algebraic means and
eqs. (\ref{subs}) become a system of four linear partial differential equations for the function
$\Delta(y,z)$. In order to integrate this system, one can proceed with the first equation which is of the Euler type.
The corresponding solution will fix the dependence of $\Delta$ on $y$ and will involve three arbitrary functions of $z$.
Their explicit form will be determined from three equations remaining in (\ref{subs}).

When constructing
an explicit solution, it should be remembered that there are nontrivial integrability conditions
associated with the system (\ref{subs}). For example, from eq. (\ref{subs}) one readily gets the
following restrictions:
\be\label{integ}
z \frac{\partial \Sigma_1}{\partial z} -y \frac{\partial \Sigma_4}{\partial y}-2\Sigma_4=0, \quad
y\frac{\partial \Sigma_2}{\partial y} -z \frac{\partial \Sigma_3}{\partial z}-2\Sigma_3=0, \quad
y \frac{\partial \Sigma_3}{\partial y} -z \frac{\partial \Sigma_4}{\partial z}+ \Sigma_3-\Sigma_4=0.
\ee
As was explained above, all the subsidiary functions are expressed in terms of $V$.
So, the integrability conditions (\ref{integ}) will be either identically satisfied, or
they will give rise to a nontrivial additional constraint on the form of the prepotential $V$.
A closed expression for such a constraint is still to be found.

Thus, we have demonstrated that for the three--body Calogero model the problem of constructing a $N=4$
superconformal extension reduces to the integration of four linear partial differential equations
(\ref{subs}).\\

\noindent
{\bf 4. Conclusion}\\

To summarize, in this paper we have constructed a set of generators which comprise symmetries of a
multi--particle mechanics in one dimension and form a $N=4$ superconformal algebra.
The WDVV equation was shown to play the central role in this framework. We treated in detail the
two--body $N=4$ superconformal Calogero model and presented some partial results on the three--body problem.

Since the connection between the WDVV equation and $N=4$ supeconformal multiparticle mechanics seems quite
intriguing, a few related problems deserve further investigation.
First of all, it would be interesting to find a topological field theory interpretation of a
generalization of the WDVV equation considered in this paper. Then, as mentioned above, the WDVV equation
may be viewed as the integrability condition for the existence of a vector field covariantly constant with
respect to the "connection" $W^{ijk}$. The geometry behind this equation remains to be clarified.

When considering the Calogero model in section 3, we found that the corresponding $N=4$ superconformal
extension is not unique. A similar observation for a $N=2$ Calogero model was recently made in
ref. \cite{gh}. It would be interesting to investigate whether the different extensions are equivalent
or related by a kind of a duality transformation. As we have seen above, the arbitrariness
is controlled by the value of the central charge in the $su(1,1|2)$ superalgebra. Then, an interesting
question is whether the central charge is fixed upon quantization.
It would be also interesting to construct multi-particle generalizations of the $N=8$ supersymmetric mechanics
models constructed recently \cite{n8}.

Finally, $N=4$ superconformal algebra examined in this work can be consistently reduced to
$N=2$ superconformal algebra. Then, one may wonder how this reduction is related to
$N=2$ superconformal Calogero model constructed by Freedman and Mende \cite{fm}.

\vspace{0.5cm}

\noindent{\bf Acknowledgements}\\

The research was supported in part by RF Presidential grants MD-252.2003.02, VNS-1252.2003.2,
INTAS grant 03-51-6346, RFBR-DFG grant 436 RYS 113/669/0-2 and RFBR grant 03-02-16193, INTAS grant No 00-0254,
European Community's Marie Curie Research Training Network contract MRTN-CT-2004-005104 and
NATO Collaborative Linkage Grant PST.CLG.979389.

\end{document}